\documentclass[aps,twocolumn,amsmath,amssymb]{revtex4}
\def\){\rangle\!\rangle}\def\({\langle\!\langle}
\def\ket#1{\vert {#1} \rangle}%
\def\set#1{{\sf #1}}
\def\transp#1{{#1}^\tau}
\def\rnk{\operatorname{rank}}
\def\Rng{\set{Rng}}\def\Ker{\set{Ker}}
\def\Supp{\set{Supp}}
\def\eigv{\operatorname{eigv}}
\def\Proof{\medskip\par\noindent{\bf Proof. 
}}
\def\qed{$\blacksquare$\par}
\newtheorem{lemma}{Lemma}
\newtheorem{theorem}{Theorem}
\begin{document}
\title{Protocols for entanglement transformations of bipartite pure states}
\author{G. Mauro D'Ariano$^{1,2}$ and  Massimiliano F. Sacchi$^1$}
\affiliation{$^1$ Unit\`a INFM, and Dipartimento di Fisica 
``A. Volta'', via A. Bassi 6, I-27100 Pavia, Italy}
\homepage{http://www.qubit.it}
\affiliation{$^2$ Department of Electrical and Computer Engineering,
Northwestern University, Evanston, IL  60208}
\date{\today} 
\begin{abstract}
We present a general theoretical framework for both deterministic and
probabilistic entanglement transformations of bipartite pure states
achieved via local operations and classical communication. This
framework unifies and greatly simplifies previous works. A necessary condition
for ``pure contraction'' transformations is given. Finally,
constructive protocols to achieve both probabilistic and deterministic
entanglement transformations are presented.
\end{abstract}
\maketitle

\section{Introduction}
The transformation of entangled states by means of local operations
and classical communication (LOCC) is a key issue
in quantum information processing \cite{Nielsen2000}, for quantum
computation \cite{spiller}, quantum teleportation \cite{BBCJPW}, and
quantum cryptography \cite{ek}. However, the detrimental effect of
losses and decoherence poses a serious problem for establishing
entangled resources at distance or in a long computing network, since
in a realistic transmission or computation the entanglement can be
considerably degraded, thus preventing, for example, successful
teleportation or dense coding. For this reason, the use of
transformations which can increase the available entanglement by means
of LOCC---although with some probability---is crucial for practical
purposes. More generally, understanding entanglement transformations
that are allowed by LOCC provides more insight in the structure and
property of nonlocality, the most prominent character of quantum
mechanics.  

\par On entanglement transformations via LOCC there are two main
results. The first is the seminal work by Nielsen \cite{nielsen99} on
deterministic transformations, which introduces majorization theory in
this context. The second is the work by Vidal \cite{vidal99}, which
addresses the more general problem of probabilistic
transformations. Such work is based on the approach of {\em
  entanglement monotones}, and gives 
conditions equivalent to weak majorization relations. The two
approaches are completely disconnected, and for practical applications
of the theory, a unified framework would be needed, especially in
consideration that the more general treatment by Vidal is more
abstract and less constructive than the Nielsen approach, which, however,
is limited only to deterministic transformations. Furthermore, from
the Vidal approach it is very difficult to recover the Nielsen treatment as a
special case, and it is quite surprising that such approach ends up as
a weak-majorization condition, without essentially using majorization
theory. This motivates a derivation of the general nondeterministic
LOCC transformations of pure states with an approach completely based
on majorization theory, generalizing the Nielsen work \cite{nielsen99}.

In this paper, we present a general framework for entanglement
transformations of bipartite pure states by means of LOCC.  In
Sec. II, we give a short and very simple proof of the theorem of Lo
and Popescu \cite{Lo-Popescu}, which is at the basis of the theory of
all LOCC, and which states that given two separate parties, say Alice
and Bob, all LOCC transformations on a pure bipartite state can be
reduced to a contraction by Alice and a unitary transformation by
Bob. We include our derivation of this theorem, since it is
particularly simple and is based on a useful technique for operator
transposition. In Sec. III, we derive the main theorem, which gives a
necessary and sufficient condition for all entanglement
transformations of pure states in terms of supermajorization
conditions, generalizing the Nielsen approach, and recovering the
result of Vidal. Here, we also provide a necessary condition for
``pure contraction'' transformations, namely those LOCC
transformations in which the target state is achieved just for a {\em
single} outcome of Alice measurement and a corresponding unitary on
Bob side. Such a condition is written in terms of submajorization relation.
In Sec. IV, explicit protocols to achieve pure, deterministic, and
probabilistic transformations are given, using a method that 
gives the Alice contraction of the LOCC in terms of the Moore-Penrose
pseudoinverse of the matrix of the entangled state.  Section V
concludes the paper summarizing the results.
\section{The Lo-Popescu theorem}
For later convenience, we introduce here the main notation used in the
paper.  Given a linear operator $O$, we denote its Hermitian conjugate
by $O^\dag $.  On a fixed basis, we write the complex conjugate and
the transpose of $O$ as $O^*$ and $\transp{O}$, respectively, so that
$O^\dag=\transp{(O^*)}$. With the notation $O^\ddag $ we denote the
Moore-Penrose inverse of $O$. We recall that the Moore-Penrose inverse
$O^\ddag$ is the unique matrix that satisfies
\begin{equation}
\begin{split}\label{MPI}
&OO^\ddag O=O\;,\qquad  O^\ddag OO^\ddag =O^\ddag \;, \\ &
OO^\ddag\quad \hbox{  and  }\quad O^\ddag O \qquad \hbox{  Hermitian}\;. 
\end{split}
\end{equation}
From Eq. (\ref{MPI}), it immediately follows that both $OO^\ddag$ and
$O^\ddag O$  are orthogonal projectors, in particular,  $OO^\ddag\equiv
P_O$ is the orthogonal projector over the range of $O$ $\Rng(O)$,
whereas $O^\ddag O\equiv P_{O^\dag}$ is the orthogonal 
projector over the support of $O$
$\Supp(O)\doteq\Ker(O)^\perp\equiv\Rng(O^\dag)$.  We write the
singular value decomposition (SVD) of $O$ as follows 
\begin{eqnarray}
O=X_O\Sigma _O Y_O \;,\label{SVDO}
\end{eqnarray}
where $\Sigma _O$ denotes the diagonal matrix whose entries are the
singular values $\sigma _i (O)$ of $O$ taken in a decreasing order, and
$X_O,Y_O$ are unitary.  The Moore-Penrose inverse $O^\ddag $ then
writes
\begin{eqnarray}
O^\ddag = Y_O^\dag \Sigma _O ^\ddag X_{O}^\dag\;,
\end{eqnarray}
where $\Sigma _O^\ddag $ is diagonal with entries $\sigma _i (O)^{-1}$
for $\sigma _i(O)\neq 0$, and zero entries for $\sigma _i(O)=0$.  

\par We remember that a quantum measurement (with discrete spectrum)
is generally described by a positive operator-valued measurement
(POVM), namely by a resolution of the identity $\sum_\lambda
M_\lambda^\dag M_\lambda =1$, where each $\lambda $ corresponds to a
possible outcome. Each operator $M_\lambda $ acts on the input states
and is necessarily a {\em contraction}, namely it satisfies $|\!|
M_\lambda |\!|\leq 1$, where $|\!|\ldots |\!|$ denotes the usual
operator norm.  \par For bipartite pure states on the Hilbert space
${\cal H}_{1} \otimes {\cal H}_{2}$ we use the following notation
\cite{bellobs}
\begin{eqnarray}
|A\)\equiv \sum_{i,j}a_{ij} |i \rangle _1\otimes |j \rangle _2   
\;,\label{kket}
\end{eqnarray}
where $\{\ket{i}_1\}$ and $\{\ket{j}_2\}$ are two chosen orthonormal
bases for ${\cal H}_{1}$ and ${\cal H}_{2}$,
respectively. Equation (\ref{kket}) introduces an isomorphism between
vectors in ${\cal H}_{1}\otimes {\cal H}_{2}$ and $n\times m$
matrices, where $n$ and $m$ are the dimensions of ${\cal H}_{1}$ and
${\cal H}_{2}$. One can easily check the relation
\begin{eqnarray}
A\otimes B |C\)=|AC\transp{B}\)\;,\label{HSrule}
\end{eqnarray}
where the transposition is defined with respect to the orthonormal
basis $\{\ket{j}_2\}$. Finally, we use the notation $A\prec B$ for
Hermitian operators $A$ and $B$ to denote the vector majorization
relation \cite{bhatia} $\eigv(A)\prec\eigv(B)$, and in the same manner
we will write $A\prec^w B$ and $A\prec_w B$ for super- and
sub-majorization, respectively.  

\par In the last part of this section, we provide a very short proof of
the following theorem \cite{Lo-Popescu}.
\begin{theorem}\label{t:LoP} All LOCC on a pure bipartite state
$|\Psi\)$ can be reduced to a contraction by Alice and a unitary
transformation by Bob. This resorts to the equivalence of any Bob
contraction $M$ with the Alice contraction $N$ assisted by Bob unitary
transformation $U$ as follows
\begin{eqnarray}
I\otimes M|\Psi\)=N\otimes U|\Psi\)\;,\label{LoP}
\end{eqnarray}
where
\begin{eqnarray}
N =K_{M\transp{\Psi }} M 
K_{\Psi } \;,\qquad 
U =K^\dag _{M\transp{\Psi }}K^\dag _{\Psi }\;,\label{due}
\end{eqnarray}
and $K_O$ is the unitary operator achieving the transposition of the
operator $O$, namely,
\begin{eqnarray}
\transp{O} = K_O O K_O^*  \;. 
\end{eqnarray}
\end{theorem}
\Proof To prove that every LOCC can be reduced to an Alice contraction
and a Bob unitary transformation, it is sufficient to prove the
equivalence (\ref{LoP}), since {\em a)} all possible elementary LOCC
in a sequence will be reduced to an Alice contraction and a Bob
unitary; {\em b)} the product of two contractions is a contraction;
{\em c)} unitary transformations are particular cases of contraction.
\par Notice that given the SVD of any linear operator $O$ as in
Eq. (\ref{SVDO}), one has
\begin{eqnarray}
\transp{O} &= &Y_\transp{O} \Sigma_O X_\transp{O}= (Y_\transp{O}
X_O^\dag)O 
(\transp{Y_O} X_O^\dag )^*
\nonumber \\&\equiv  & K_O O K_O^*  
\;,\label{ko}
\end{eqnarray}
with $K_O=Y_\transp{O} X_O^\dag$. Using Eq. (\ref{HSrule}),
Eq. (\ref{LoP}) rewrites as follows
\begin{equation}
\Psi\transp{M}=N\Psi\transp{U}.
\end{equation}
Then, from Eq. (\ref{ko}) one has
\begin{eqnarray}
\Psi \transp{M} &=& \transp{(M\transp{\Psi })}= K_{M\transp{\Psi }}
(M\transp{\Psi }) K^*_{M\transp{\Psi }}\nonumber \\& =&K_{M\transp{\Psi }}M 
K_{\Psi }\Psi K _\Psi ^* K^*_{M\transp{\Psi }}\;, 
\end{eqnarray}
which is just Eq. (\ref{LoP}) with  $N $ and $U$ given as 
in Eq. (\ref{due}).  \hfill\qed 
\section{LOCC transformations for pure states}\label{s:dentang}
In this section, we will use the following useful lemmas.
\begin{lemma}\label{l:2}
If $x\prec^w y $, then for some $v\quad  x\prec v $ and $v\geq y$. 
\end{lemma}
\Proof 
If $x\prec^w y $ one has for $2\le l\le N$
\begin{equation}
\sum_{i=l}^N x_i \geq \sum_{i=l}^N y_i
\end{equation}
and 
\begin{equation}
q\equiv \sum_{i=1}^N x_i > \sum_{i=1}^N y_i \equiv p. 
\end{equation}
Upon defining 
\begin{eqnarray}
v=(q-p+y_1,y_2,...,y_N)\; 
\end{eqnarray}
clearly one has $v\geq y$, and 
\begin{eqnarray}
\sum_{i=1}^l x_i 
&=&q- \sum_{i=l+1}^N x_i \leq q -\sum_{i=l+1}^N y_i \nonumber \\&= & 
q-p +\sum_{i=1}^l y_i =\sum_{i=1}^l v_i \;,\qquad \forall \ l 
\end{eqnarray}
namely $x \prec v$. 
\hfill\qed
\begin{lemma}\label{l:1}
If for some  $u\ \ x\geq u $ and $u\prec y$, then $x\prec^w y $. 
\end{lemma}
\Proof One has 
\begin{eqnarray}
\sum_{i=l}^N x_i &\geq &\sum_{i=l}^N u_i =\sum_{i=1}^N u_i
-\sum_{i=1}^{l-1} u_i 
\nonumber \\&  \geq & \sum_{i=1}^N y_i
-\sum_{i=1}^{l-1} y_i =\sum_{i=l}^N y_i
\;,\qquad \forall \ l 
\end{eqnarray}
namely, $x\prec^w y $. 
\hfill\qed 
We notice that both the above lemmas hold also in the reverse
direction (for a proof see Ref. \cite{Marshall}, pp. 11 and 123).  
\par Moreover, we will make extensive use of the following
theorem.
\begin{theorem}[Uhlmann]\label{t:Uhlmaj} 
For Hermitian operators $A$ and $B$, one has
$A\prec B$ if and only if there are probabilities $q_\lambda $ 
and unitary operators $W_\lambda $ such that 
\begin{equation}
A=\sum_\lambda  q_\lambda W^\dag _\lambda  B W_\lambda\;. 
\end{equation}
\end{theorem}
\Proof See Ref. \cite{Nielsen2000}, p. 575.  

\par\noindent Theorem \ref{t:Uhlmaj} relates majorization between
Hermitian operators with a particular form of completely positive
maps, namely those achievable through a random unitary evolution. In
the terminology of quantum-information channels, such maps correspond
to external-random-field channels that are a subclass of bistochastic
channels (which send the identity operator into itself).  For a qubit
system (${\cal H }={\cal C}^2$), the set of bistochastic and
external-random-field channels coincide \cite{otto}.  

\par In the following, we derive the necessary and sufficient condition
for all entanglement transformations of pure states in terms of
supermajorization conditions. The theorem generalizes Nielsen
approach \cite{nielsen99} and recovers the result of Vidal
probabilistic transformations \cite{vidal99}. Moreover, we provide a 
necessary condition for ``pure contraction'' transformations,
namely, those LOCC transformations in which the target state is
achieved just for a {\em single} outcome of Alice measurement and a
corresponding unitary on Bob side. In the proof a relevant role is
played by the intermediate state (denoted in the following by $|Q
\rangle\!\rangle $). In a transformation from $|A \rangle\!\rangle $
to $|B \rangle\!\rangle $, the state $|Q \rangle\!\rangle $ will be
reached from $|A \rangle\!\rangle $ deterministically, whereas the
final probabilistic transformation $|Q \rangle\!\rangle \to |B
\rangle\!\rangle $ will be obtained through a pure contraction (for
deterministic transformations one has $Q\equiv B$).
 \par We are now ready to prove the following
\begin{theorem}
The state transformation $|A\)\to|B\)$ is possible by LOCC iff 
\begin{equation}
AA^\dag\prec^w pBB^\dag \;,\label{MajBA}
\end{equation}
where $p\le 1$ is the probability of achieving the
transformation. 
\par\noindent A necessary condition to be satisfied is
$\rnk(A)\ge\rnk(B)$. \par\noindent 
In particular, the transformation is deterministic
($p=1$) iff $AA^\dag\prec BB^\dag$.  
\par\noindent Finally, if there is a {\em pure} LOCC
that achieves the state transformation with probability $p$, we must have
\begin{equation}
pBB^\dag\prec_w AA^\dag.\label{MajBA2}
\end{equation}
\end{theorem}
\Proof Assume that $AA^\dag\prec^w pBB^\dag$. 
From Lemma \ref{l:2} there is an operator $Q$ with
$\Sigma_Q^2 \geq p\Sigma_B^2$ and $AA^\dag \prec QQ^\dag $. The state 
$|Q \rangle\!\rangle $ represents the intermediate state that can be
achieved deterministically from $|A \rangle\!\rangle  $. In fact, 
Uhlmann Theorem \ref{t:Uhlmaj} guarantees the existence of
a set of unitaries $W_\lambda$ and probabilities $q_\lambda$ such that   
\begin{equation}
AA^\dag =\sum_\lambda q_\lambda W_\lambda^\dag QQ^\dag W_\lambda
 \;.\label{ranW}
\end{equation} 
Now, define the Alice measurement $\{M_\lambda\} $ such that 
\begin{eqnarray}
M_\lambda \sqrt{AA^\dag}=\sqrt{q_\lambda QQ^\dag}W_\lambda
\;, 
\end{eqnarray}
namely, 
\begin{equation}
M_\lambda AA^\dag M_\lambda^\dag=q_\lambda QQ^\dag\;,\quad\forall
\lambda\;.  
\end{equation}
We can always choose $M_\lambda $ such that $\Supp({M_\lambda
})\subseteq \Rng(A)$, and show that 
$M_\lambda $ is a contraction, since
\begin{eqnarray}
\sqrt{AA^\dag }\sum_\lambda M^\dag _\lambda M_\lambda 
\sqrt{AA^\dag }= A A^\dag 
\;, 
\end{eqnarray}
and so $\sum_\lambda M_\lambda
^\dag M_\lambda  =P_A\leq I$. 
Then, there exists a set of unitary operators
$U_\lambda$ such that $M_ \lambda A\transp{U}_\lambda =\sqrt{q_\lambda }Q$, namely,
\begin{equation}
M_\lambda\otimes U_\lambda |A\)=\sqrt{q_\lambda} |Q\)\;,
\end{equation}
so that the transformation from $|A\)\to|Q\)$ can be achieved
deterministically. Now, since $\Sigma_Q^2\geq p\Sigma_B^2$ 
one can define the contraction 
\begin{equation}
\tilde 
N=\sqrt p \sum_l \frac {\sigma _l(B)}{\sigma _l (Q)}|l \rangle \langle l|
\;\label{facile}
\end{equation}
so that 
\begin{eqnarray}
\tilde N\Sigma _Q^2 \tilde N^\dag =p\Sigma _B^2\;.\label{NAN}
\end{eqnarray}
By using the SVD of both $Q$ and $B$, Eq. (\ref{NAN}) rewrites
as follows
\begin{equation}
\tilde N X_ Q^\dag QQ^\dag X_Q \tilde N^\dag =p X _B^\dag BB^\dag X_B\;,
\end{equation}
which means that there exists a unitary transformation $V$ such that
\begin{equation}
 NQ\transp{V}\label{NAV}=\sqrt{p}B\;,
\end{equation}
where $N= X_B\tilde N X_Q^\dag$. Equation (\ref{NAV}) is equivalent to the
entanglement transformation
\begin{equation}
N\otimes V|Q\)=\sqrt{p}|B\)\;,
\end{equation}
namely, there is a pure LOCC occurring with probability $p$, which transforms
the state $|Q\)$ into  $|B\)$. As we have seen before,
the transformation $|A\)\to|Q\)$ can be achieved deterministically,
whence we conclude that $|A\)\to|B\)$ can be
achieved with probability $p$, namely the statement of the first part of
the theorem.
\par Reversely, assume that the transformation $|A \rangle \!\rangle
\to |B \rangle \!\rangle $ can be achieved via LOCC with probability
$p$. According to Theorem \ref{t:LoP} every LOCC is equivalent to a
measurement performed by Alice followed by a conditional unitary 
by Bob. Therefore, the joint Alice-Bob state
$R$ will evolve as follows:
\begin{eqnarray}
R\to \sum_\lambda M_\lambda\otimes U_\lambda R
M_\lambda^\dag\otimes U_\lambda^\dag\;,
\end{eqnarray}
where
\begin{eqnarray}
\sum_\lambda M_\lambda^\dag M_\lambda=I\;.\label{completeM}
\end{eqnarray}
If the state $|A\)$ goes to $|B\)$, we must
have for a subset $\set{S}$ of the possible outcomes $\lambda $
\begin{equation}
M_\lambda\otimes U_\lambda
|A\)=\sqrt{p_\lambda}|B\)\;,\qquad\forall \lambda \in \set{S}\;,\label{Niels}
\end{equation}
where $\sum_{\lambda \in S} p_\lambda =p$ denotes the overall
probability of the transformation $|A\)\to|B\)$.  From
Eq. (\ref{Niels}), we need to have
\begin{equation}
M_\lambda  A\transp{U}_\lambda =\sqrt{p_\lambda }B\;,\quad \forall \ \lambda
\in\set{S},\label{MAU}
\end{equation}
and, therefore, each probability $p_\lambda$ is given by
\begin{equation}
|\!|M_\lambda A\transp{U}_\lambda |\!|_2^2=|\!|M_\lambda
A|\!|_2^2=p_\lambda \;,
\end{equation}
where $|\!|O|\!|_2=\sqrt{\hbox{Tr}[O^\dag O]}$ denotes the usual Frobenius norm. 
The condition (\ref{MAU}) can be satisfied only if
$\rnk(A)\ge\rnk(B)$, 
i.e. we can only decrease the Schmidt number of the entangled state.  
\par From Eq. (\ref{Niels}), one has $\forall\ \lambda \in\set{S}$
\begin{align}
M_\lambda AA^\dag M_\lambda^\dag=&p_\lambda BB^\dag\;,\\
\intertext{namely, by polar decomposition}
M_\lambda\sqrt{AA^\dag}=&\sqrt{M_\lambda AA^\dag
M_\lambda^\dag}V_\lambda =\sqrt{p_\lambda BB^\dag}V_\lambda\;.\label{polV}
\end{align}
From Eq. (\ref{completeM}), we have
\begin{eqnarray}
\frac 1p AA^\dag&=&\frac 1p \sqrt{AA^\dag}\sum_\lambda M_\lambda^\dag
M_\lambda\sqrt{AA^\dag}\nonumber \\&  \geq &\frac 1 p  
\sqrt{AA^\dag}\sum_{\lambda\in\set{S}} M_\lambda^\dag
M_\lambda\sqrt{AA^\dag}\;,
\end{eqnarray}
and using Eq. (\ref{polV}), we obtain
\begin{equation}
\frac 1p AA^\dag\ge 
\sum_{\lambda\in\set{S}} \frac {p_\lambda} {p} V_\lambda^\dag BB^\dag V_\lambda
\equiv QQ^\dag \;.\label{ranU}
\end{equation}
By Uhlmann Theorem \ref{t:Uhlmaj}, we have 
\begin{equation}
QQ^\dag \prec BB^\dag,\label{froml2}
\end{equation}
and from Lemma \ref{l:1} we get the statement, namely, 
$AA^\dag\prec^w pBB^\dag$.

\par We now prove the last part of the theorem, regarding the pure
LOCC case. In a pure contraction transformation the target state is
achieved just for a {\em single} outcome of Alice measurement and a
unitary performed by Bob. Such a
transformation that occurs with probability $p$ is given by
\begin{equation}
M\otimes U|A\)=\sqrt{p}|B\)\;,\label{LOCC}
\end{equation}
and we need to have 
\begin{align}
MA\transp{U}=&\sqrt{p}B\;,\label{MAUagain}\\ 
|\!|MA\transp{U}|\!|_2^2=&|\!|MA|\!|_2^2=p\;.
\end{align}
Again, this is possible if $\rnk(A)\ge\rnk(B)$. 
Using the SVD of $A$ and $B$ as in Eq. (\ref{SVDO}), Eq. (\ref{MAUagain})
rewrites in terms of the diagonal matrices $\Sigma_{A}$ and  $\Sigma_{B}$ 
as follows
\begin{equation}
\tilde{M}\Sigma_A\tilde{U}=\sqrt{p}\,\Sigma_B\;,\;\label{MAU1}
\end{equation}
with
\begin{equation}
\tilde{M}=X_B^\dag MX_A\;,\qquad\tilde{U}=Y_A\transp{U}Y_B^\dag\;.\label{MAU2}
\end{equation}
Equation (\ref{MAU1}) leads to
\begin{equation}
\tilde{M}\Sigma_A^2\tilde{M}^\dag=p\Sigma_B^2\;,
\end{equation}
namely,
\begin{equation}
\sum_k S_{kl}\sigma_k^2(A)=p\sigma_l^2(B)\;,\label{S}
\end{equation}
where $S_{kl}\doteq|\langle  l|\tilde{M}|k\rangle |^2$ is a substochastic
matrix \cite{Marshall}, since 
\begin{alignat}{2}
\sum_k S_{kl}=&\langle  l|\tilde{M}\tilde{M}^\dag|l\rangle \le
&|\!|M^\dag|\!|^2\le& 1\;,\\
\sum_l S_{kl}=&\langle  k|\tilde{M}^\dag\tilde{M}|k\rangle \le&|\!|M|\!|^2\le& 1\;.
\end{alignat}
Since Eq. (\ref{S}) with $S$ substochastic is equivalent
\cite{Marshall} to $p\sigma^2(B)\prec_w \sigma^2(A)$, namely
$pBB^\dag\prec_w AA^\dag$, we have proved that Eq. (\ref{MajBA2}) is a
necessary condition for the LOCC transformation (\ref{LOCC}), namely,
the last statement of the theorem.  \hfill\qed

\section{Explicit protocols}
\subsection{Pure transformation}
Pure LOCC transformations are achieved by a single contraction on
Alice side, assisted by a unitary by Bob. These are the most general
one-way LOCC operations. In this case, we have the following theorem.
\begin{theorem}
The transformation $|A\) \rightarrow |B\)$ can be achieved with
probability $p$ by a pure LOCC transformation 
iff one can find a unitary operator $U$
and linear operator $N$ such that 
\begin{equation}
M=\sqrt{p}BU^*A^\ddag+N(I-AA^\ddag)\label{Mparam}
\end{equation}
is a contraction. The transformation is then obtained as 
\begin{equation}
M\otimes U|A\)=\sqrt{p}|B\)\;\label{mua}
\end{equation}
\end{theorem}
\Proof Notice that Eq. (\ref{mua})  is equivalent to
the identity
\begin{equation}
MA=\sqrt{p}BU^*\;.
\end{equation}
Since both sides of the identity must have the same kernel it follows
that
\begin{equation}
MA=\sqrt{p}BU^*A^\ddag A\;,
\end{equation}
and, multiplying both sides by $A^\ddag$, we have
\begin{equation}
MP_A=\sqrt{p}BU^*A^\ddag AA^\ddag=\sqrt{p}BU^*A^\ddag P_A 
=\sqrt{p}BU^*A^\ddag\;.
\end{equation}
The general solution of the last equation is
\begin{equation}
M=\sqrt{p}BU^*A^\ddag P_A+N(I-P_A)\;\label{gene}
\end{equation}
with arbitrary $N$ and, in fact, one can easily check that 
\begin{equation}
MA=\sqrt{p}BU^*A^\ddag A=\sqrt{p}BU^*.
\end{equation}
The unitary $U$ and the operator $N$ should be taken such that
$M$ is a contraction. This is not always possible. However, a
sufficient condition is  
\begin{eqnarray}
p\Sigma _B^2\leq \Sigma _A^2\;.\label{pba} 
\end{eqnarray}
In fact, by taking $N=0$ and $U=Y_B^T Y_A^*$, one has $M={\sqrt p}
X_B\Sigma _B \Sigma ^\ddag _A X_A^\dag  $, and then, for
Eq. (\ref{pba}), $|\!|M |\!|=\sqrt p \,|\!|\Sigma _B \Sigma _A^\ddag
|\!|\leq 1$.
\hfill\qed
\subsection{Deterministic transformation}
In the entanglement transformations, the first
part of the protocol is a deterministic transformation from $|A
\)$ to $|Q\)$. The majorization relation $AA^\dag \prec QQ^\dag  $
implies Theorem \ref{t:Uhlmaj}, namely the existence of
a set of unitaries $W_\lambda$ and probabilities $q_\lambda$ such that   
\begin{equation}
AA^\dag =\sum_\lambda q_\lambda W_\lambda^\dag QQ^\dag W_\lambda
 \;.\label{ranW2}
\end{equation} 
In order to construct explicitly the protocol, one needs to find 
contractions $M_\lambda $ and unitaries $U_ \lambda $
versus the unitary operators $W_\lambda $ that appear in
Eq. (\ref{ranW2}) such that one has
\begin{eqnarray}
M_\lambda\otimes U_\lambda |A\)=\sqrt{q_\lambda} |Q\)\;.
\;\label{mza}
\end{eqnarray}
We have seen that the general form of the solution of Eq. (\ref{mza}) 
is given by
\begin{eqnarray}
M _\lambda =\sqrt{q_\lambda } QU_\lambda ^* A^\ddag +
N_\lambda (1- AA^\ddag )\;.\label{ml}
\end{eqnarray}
We have now the following theorem
\begin{theorem}
In Eq. (\ref{ml}), we can always take 
\begin{eqnarray}
N_\lambda =0\;,\qquad U_\lambda ^* =Y_Q^\dag X_Q^\dag W_\lambda X_A Y_A
\;,\label{zl}
\end{eqnarray}
where $X_O,Y_O$ are the operators defined in Eq. (\ref{SVDO}) such 
that $M_\lambda $ is a contraction and Eq. (\ref{mza}) is satisfied.
\end{theorem}
\Proof Substituting Eq. (\ref{zl}) in Eq. (\ref{ml}), one has
\begin{eqnarray}
&&\sum _\lambda {M^\dag _\lambda M_\lambda }=
\sum_\lambda q_\lambda (A^\ddag )^\dag \transp{U_\lambda } Q^\dag 
QU_\lambda ^* A^\ddag  \nonumber \\&  &=
\sum_\lambda q_\lambda (A^\ddag )^\dag Y_A^\dag X_A ^\dag W^\dag _\lambda 
X_Q Y_Q  Q^\dag 
Q  Y_Q^\dag X_Q^\dag W_\lambda X_A Y_A A^\ddag  \nonumber \\& &= 
\sum_\lambda q_\lambda (A^\ddag )^\dag Y_A^\dag X_A ^\dag W^\dag _\lambda 
Q Q^\dag  W_\lambda X_A Y_A A^\ddag \;, 
\end{eqnarray}
and using Eq. (\ref{ranW2}), one has
\begin{eqnarray}
&&\sum _\lambda {M^\dag _\lambda M_\lambda }=
(A^\ddag )^\dag Y_A^\dag X_A ^\dag A A^\dag X_A Y_A A^\ddag
  \\& & =
(A^\ddag )^\dag  A^\dag A A^\ddag   =
(A A^\ddag )^\dag  A A^\ddag   = A A^\ddag   = P_A \leq I 
\;.\nonumber  
\end{eqnarray}
Hence, $M_\lambda $ are contractions. 
The completeness of the measurement can be guaranteed by the further
contraction
\begin{eqnarray}
M_0=V (I -A A^\ddag )\;, 
\end{eqnarray}
where $V$ is an arbitrary unitary operator.
\par For any outcome $\lambda $ on Alice side, Bob performs the unitary
$U_\lambda $. 
Using Eqs. (\ref{ml}) and (\ref{zl}), one
has 
\begin{eqnarray}
M_\lambda A \transp{U_\lambda }&=&\sqrt {q_\lambda }
Q  Y_Q^\dag X_Q^\dag W_\lambda X_A Y_A A^\ddag   
A Y_A^\dag X_A ^\dag W^\dag _\lambda 
X_Q Y_Q \nonumber \\ &=&\sqrt {q_\lambda }
Q  Y_Q^\dag X_Q^\dag W_\lambda A A^\ddag  W^\dag _\lambda 
X_Q Y_Q  \;.\label{maz}
\end{eqnarray}
From Eq. (\ref{ranW2}), it follows that $\Rng(W_\lambda ^\dag 
 QQ^\dag W_\lambda )\subseteq \Rng(AA^\dag)\equiv\Rng(A)$, namely, 
\begin{eqnarray}
P_A=P_{AA^\dag}\geq 
W_\lambda ^\dag P_{ QQ^\dag }W_\lambda \;,\qquad \forall \lambda \;.\label{pa}
\end{eqnarray}
Hence,  by multiplying  both sides  of Eq. (\ref{pa}) on the left by 
$Y_Q^\dag X_Q^\dag W_\lambda$ and on the right by $W^\dag _\lambda 
X_Q Y_Q $, one obtains  
\begin{eqnarray}
 Y_Q^\dag X_Q^\dag W_\lambda P_A  W^\dag _\lambda 
X_Q Y_Q  &\geq & 
 Y_Q^\dag X_Q^\dag P_{QQ^\dag }
X_Q Y_Q  \nonumber \\&= & P_{Q^\dag Q} =P_{Q^\dag }
\;. 
\end{eqnarray}
The projector on $\Rng(Q^\dag )$ coincides with the
projector on $\Ker(Q)^\perp\equiv\Supp(Q)$. So
Eq. (\ref{maz}) gives 
\begin{eqnarray}
M_\lambda A \transp{U_\lambda }&=&\sqrt {q_\lambda }Q\;, 
\end{eqnarray}
which is equivalent to Eq. (\ref{mza}).  
\hfill\qed 

\par According to our derivation, given explicitly Eq. (\ref{ranW2}),
one can perform the contractions $M_\lambda $ and the unitaries
$U_\lambda $ to achieve the transformation $|A\)\to |Q\)$. The problem
of looking for a POVM with minimum number of outcomes (thus,
minimizing the amount of classical information sent to Bob) is reduced
to find the transformation (\ref{ranW2}) with minimum number of
unitaries.  \par One can resort to a constructive algorithm to find a
bistochastic matrix $D$ that relates the vectors $\sigma ^2 (A)$ and
$\sigma ^2 (Q)$ of the singular values of $A$ and $Q$, namely
\begin{eqnarray}
\vec \sigma ^2 (A)= D \vec \sigma ^2 (Q) \;. 
\end{eqnarray}
Then Birkhoff theorem \cite{Marshall} allows to write $D$ as a convex
combination of permutation matrices 
\begin{eqnarray}
D=\sum _\lambda q_\lambda \Pi _\lambda  \;. 
\end{eqnarray}
In terms of $\Sigma _A$ and $\Sigma_Q$ one 
has 
\begin{eqnarray}
\Sigma_A^2 =\sum_\lambda q_\lambda \Pi ^\dag _\lambda 
\Sigma_Q^2 \Pi  _\lambda\;,\label{pil}
\end{eqnarray}
where $\Pi _\lambda =\sum _l |l \rangle \langle \Pi _\lambda (l)|$.
In this way one obtains Eq. (\ref{ranW2}), with $W_\lambda =X_Q \Pi
_\lambda X_A^\dag $. Using the corresponding expressions for the
contractions $M_\lambda $ and unitaries $U_ \lambda $ one recovers the
result of Ref. \cite{schack}.  
Notice that Caratheodory's theorem always allows to reduce the number of
permutations in Eq. (\ref{pil}) to $(d-1)^2+1$, for $d$-dimensional
Alice Hilbert space.    
\subsection{Probabilistic transformation}
\par The second part of the protocol, namely, the contraction that 
provides the state $|B\)$ from $|Q \)$, is needed only for
probabilistic transformations.  It is a pure contraction given by
\begin{eqnarray}
N \otimes V |Q\)=|N Q \transp{V} \)=
\sqrt p |B\)\; 
\end{eqnarray}
with 
\begin{eqnarray}
N=\sqrt p X_B \Sigma _B \Sigma _Q^\ddag X_Q^\dag \; 
\end{eqnarray}
and 
\begin{eqnarray}
\transp{V}=Y_Q^\dag Y_B\;. 
\end{eqnarray}
In fact for Lemma \ref{l:2}, one has $\Sigma _Q^2 \geq p \Sigma _B^2$,
which implies that 
$\Sigma _B \Sigma _Q^\ddag \Sigma _Q =\Sigma _B $. Then 
\begin{eqnarray}
N Q \transp{V} &=&\sqrt p X_B \Sigma _B \Sigma _Q ^\ddag \Sigma _Q Y_B
\nonumber \\& =& \sqrt p X_B \Sigma _B  Y_B = \sqrt p B
\;.
\end{eqnarray}
\section{Conclusions}
In this paper, we presented a general theoretical framework for both
deterministic and probabilistic entanglement transformations of
bipartite pure states achieved via LOCC transformations. We have
generalized Nielsen work based on majorization theory \cite{nielsen99}
in order to include the more general results by Vidal \cite{vidal99},
which were based on the approach of the entanglement monotones.  The
main theorem gives an if and only if condition for all entanglement
transformations of pure states in terms of super-majorization
conditions. We also provided a  necessary submajorization
condition for pure transformations, which allows to write each
contraction in terms of the Moore-Penrose pseudoinverse of the matrix
of the entangled state. This led to explicit protocols to
achieve pure, deterministic, and probabilistic LOCC. 

\par We notice that all theorems have been derived in finite
dimensions, but they can be easily extended to infinite dimensions for
contractions that are compact operators and for normalized entangled
states corresponding to Hilbert-Schmidt operators. Thus, our results
also apply to continuous variables.

\section*{Acknowledgments}
We are grateful to Shashank Virmani for useful discussions.
This work has been jointly founded by the EC under the program
EQUIP (Contract No. IST-1999-11053)  and 
ATESIT (Contract No. IST-2000-29681). G. M. D. acknowledges support
from the Department of Defense Multidisciplinary University Research
Initiative (MURI) program administered by the Army Research Office
under Grant No. DAAD19-00-1-0177.

\end{document}